\documentclass[12pt,prb]{revtex4}

\usepackage[T1]{fontenc}
\usepackage[latin9]{inputenc}
\usepackage[a4paper]{geometry}
\usepackage{color}
\usepackage{amsmath}
\usepackage{amssymb}
\usepackage{graphicx}

\usepackage{graphicx}
\usepackage{epsfig}
\usepackage{amsbsy}
\usepackage{amsmath}
\usepackage{color}
\usepackage{epsf}
\usepackage{multirow}

\usepackage{setspace}

\begin{document}
\title{Spin-orbit interaction in bent carbon nanotubes: resonant spin transitions}

\author{E. N. Osika and B. Szafran}
 \address{AGH University of Science and Technology, \\ Faculty of Physics and Applied Computer Science,
al. Mickiewicza 30, 30-059 Krak\'ow, Poland}

\begin{abstract}
We develop an effective tight-binding Hamiltonian for spin-orbit (SO) interaction in bent carbon nanotubes (CNT) for
the electrons forming the $\pi$ bonds between the nearest neighbor atoms.
We account for the bend of the CNT and the intrinsic spin-orbit interaction which introduce mixing of $\pi$ and $\sigma$ bonds between the $p_z$ orbitals along the CNT.
The effect contributes to the main origin of the SO coupling -- the folding of the graphene plane into the nanotube.
We discuss the bend-related contribution of the SO coupling for resonant single-electron spin and charge transitions
in a double quantum dot.
We report that although the effect of the bend-related SO coupling is weak for the energy spectra,
it produces a pronounced increase of the spin transition rates driven by an external electric field.
We find that spin-flipping transitions driven by alternate electric fields have usually larger rates when accompanied by charge shift from one dot to the other.
Spin-flipping transition rates are non-monotonic functions of the driving amplitude since they are masked by stronger spin-conserving charge transitions.
We demonstrate that the fractional resonances -- counterparts of multiphoton
transitions for atoms in strong laser fields -- occurring in electrically controlled nanodevices already at moderate ac amplitudes  --
can be used to maintain the spin-flip transitions.
\end{abstract}

\maketitle

\section{Introduction}
The spins of electrons confined in quantum dots are considered candidates for the quantum information processing \cite{divi},
and the external control of confined spins has attracted a lot of attention \cite{hanson}.
 In III-V devices the possible applicability is limited by low coherence time due to the interaction of the electrons with the nuclear spin field \cite{tar,c13}. The absence of the nuclear spin-field in carbon-based materials: graphene \cite{graphene} and carbon nanotubes \cite{cnt,cnt2} makes them attractive for coherent spin manipulation.
In contrast to graphene\cite{graphene} in which an electrostatic confinement of charge
carriers is excluded by the Klein tunneling \cite{kt}, formation of quantum dots by external
potentials is straightforward in semiconducting carbon nanotubes (CNTs) \cite{cnt}.
The electron spin can be controlled with electric fields by the spin-orbit coupling. The spin-orbit coupling
 in graphene is of an atomic origin and couples the $p_z$ orbitals (forming the $\pi$ bonds) and the in-plane $p_x/p_y$ which
 form the $\sigma$ between the ions. However, due to the orthogonality of $p_z$ and in-plane orbitals the spin-orbit interaction does not mix the electron spins in the band structure \cite{cnt2}. This is no longer
the case once the graphene plane is folded into the CNT\cite{ando,bulaev,soc2,delvalle},
for which the effect of spin-orbit coupling are clearly observed in spite of low atomic number of carbon \cite{Ku,jespersen,large,taco,ffpei,pecker}.

The spin control in CNTs is intensely studied by both experiment \cite{Ku,jespersen,large,taco,ffpei,pecker}
and theory \cite{ando,bulaev,soc2,delvalle,klino1,pal,flensberg,palprl}.  Recently, experimental observations
of spin transitions driven by external ac electric field were reported \cite{taco,ffpei} with
the electric-dipole spin resonance (EDSR) mechanism that was applied previously for III-V quantum dots \cite{edsrIIIV,fractional}.
The effects of the EDSR-driven transitions \cite{taco,ffpei,edsrIIIV,fractional,fracto,eosprb}
are observed by lifting the blockade of the current flow through a biased double quantum dot.

The spin-transitions in carbon nanotubes can be triggered by a bend of the CNT \cite{flensberg}, which induces
a dependence of the Zeeman splitting on electron position within the nanotube via the strong anisotropy of the effective Land\'e factor $g$.
A subsequent experiment \cite{taco} indicated that the bend of the CNT has indeed a dominant contribution to the EDSR effect.
Recent papers \cite{eosprb,pei,li} investigated the CNT bend in the context of
the symmetry of single-electron wave functions in external electric fields \cite{eosprb}, atomic disorder \cite{li},
and the leakage current \cite{pei}.

In this paper we consider the contribution of the bend of the CNT to the principal spin-orbit interaction resulting from folding the graphene plane into the tube \cite{ando}.  We develop an effective tight-binding Hamiltonian for the SO coupling resulting from the bend of the CNT as a whole. We consider the effects
 resulting from the bend of the CNT for electron spin manipulation in a double quantum dot.
 The contribution of bend-related SO interaction for the results should be expected small since the radii of the bend are by at least an order of magnitude larger than the CNT radii.
We indeed find that the effects of the spin-orbit coupling resulting from the bend for the energy spectra are negligible,
but the spin transition driven by EDSR are accelerated several times.
We find that generally the spin-flip transitions occur with a larger rate when the interdot charge
transfer accompanies the inversion of the spin as compared to intradot spin flips with fixed charge
distribution.  The resonant frequencies for transitions involving charge transfer with conserved
or flipped spin are close also at high magnetic field. We find that the direct Rabi oscillation
with spin inversion becomes attenuated by the spin-conserving interdot charge transition for a larger amplitude of the AC field.
However, the spin-flip transitions can still be reached by the fractional resonances.

The experimental conditions with ac  bias applied to the CNT resemble the ones present
for atoms and molecules in strong laser fields. In these systems besides the common Rabi oscillation,
fractional resonances -- also known as higher harmonics generation -- are observed \cite{ffpei,fractional}, which are counterparts of  multiphoton transitions of quantum optics \cite{lewenstein}.
We indicate that in conditions when the transition
with spin flip and interdot charge transfer is masked by a strong spin-conserving transition,
one can employ fractional resonances to perform the spin-flips.
We indicate that the rates of spin-flipping transitions are non-monotonic functions
of the driving frequency due to overlap with the strong spin-conserving resonances.
The SO coupling due to the bend largely increases the width of the fractional resonances for spin-flipping transitions.

\begin{figure*}[htbp]

\includegraphics[width=0.9\textwidth]{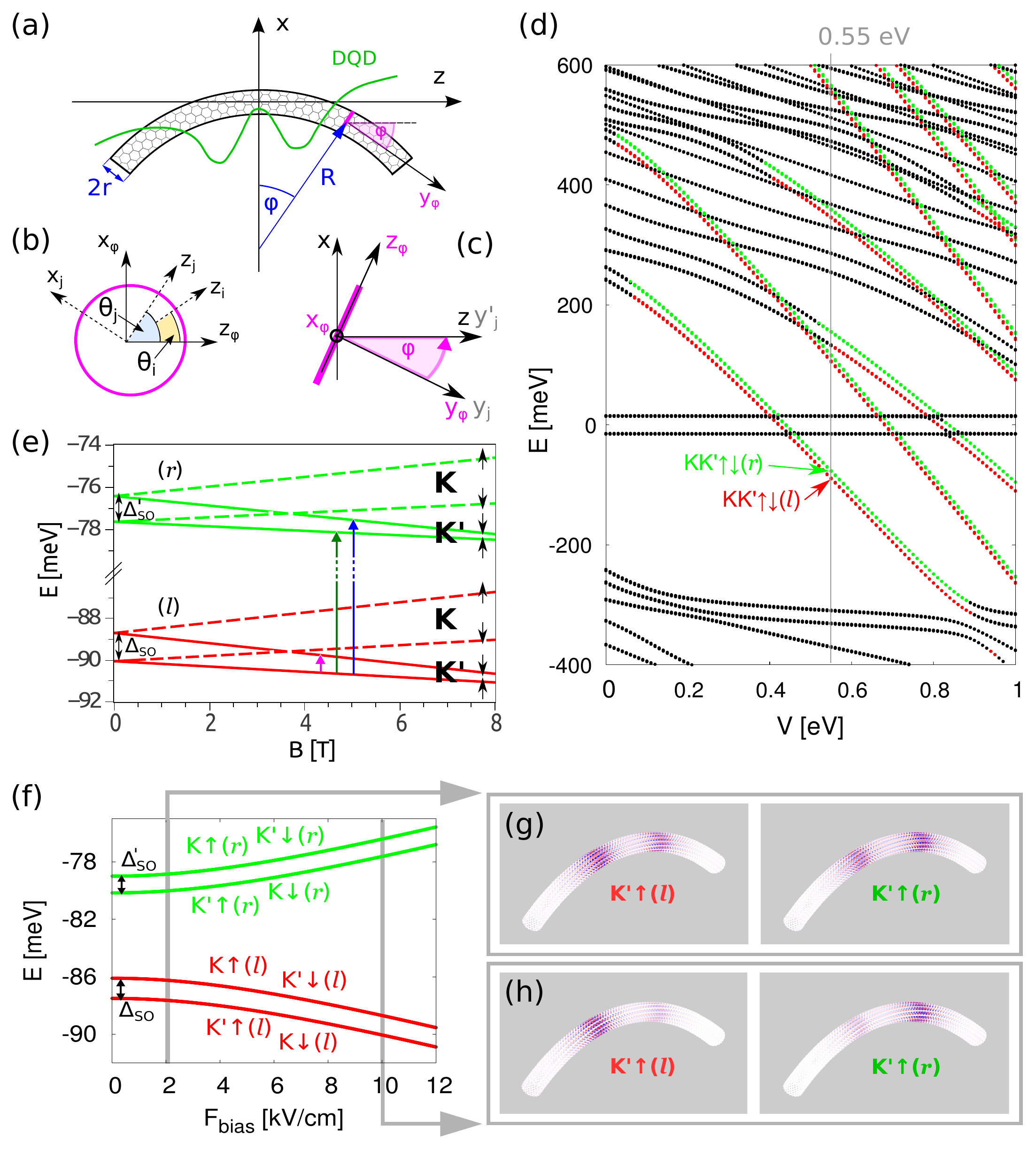}
\caption{\footnotesize \linespread{0.5} (a) Schematics of the considered system of a nanotube of length $L=31.8$ nm (150 elementary cells),
radius $r=0.78$ nm, with a circular bend of an arc radius $R=30$ nm.
Magnetic field is applied in the $z$ direction. Along the $z$ direction a double quantum dot is defined and a bias applied.
(b) Cross section of the pink ring of (a) with the definition of angles $\theta$ used for hopping parameters,
and a local coordinate system with $z_i$ along the $p_z$ orbital of $i$th ion. (c) An enlarged fragment of (a)
and the coordinate system $x_\phi,y_\phi,z_\phi$ used to rotate the spins defined along the $z_j$ direction.
(d) Energy spectrum for $B=0$ and bias field $F_{bias}=10$ kV/cm in function of the depth of the Gaussian quantum dots. The red and green
lines correspond to energy levels which are localized mostly in the left and right dots respectively. Each of the energy levels plotted with green and red lines is nearly fourfold degenerate
with respect to the spin and valley.
(e) Energy levels as functions of the magnetic field at $F_{bias}=10$ kV/cm. The purple, green and blue arrows show the allowed, i.e. valley conserving transitions from the $K'\uparrow(l)$ ground state
to $K'\downarrow(l)$, $K'\uparrow(r)$ and $K'\downarrow(r)$ energy levels, where $l$ and $r$ denote the localization of the most of the charge in the left or the right quantum dot.
(f) Energy levels for $V=0.55$ eV as functions of the bias for $B=0$,
and charge density for the branches localized in the left and right dot at $F_{bias}=2$ kV/cm (g) and $F_{bias}=10$ kV/cm (h).} \label{1sc}
\end{figure*}

\section{Theory}

We consider a CNT of length $L=31.8$ nm  -- see Fig. \ref{1sc}(a).  The CNTs geometry is defined by a chiral vector \cite{ando}
${\bf C}_h=n_1{\bf a}_1+n_2{\bf a}_2$  with primitive lattice vectors, ${\bf a}_1=a_0(1,0)$ and ${\bf a}_2=a_0(1/2,\sqrt{3}/2)$ and $a_0=0.246$ nm.
Here we consider a zigzag nanotube with $n_1=20$, $n_2=0$ for which the CNT radius is $r=0.78$ nm \cite{jang}, and for which an energy gap is
present in the dispersion relation that allows for confinement of charge carriers by external potentials.
We assume that the radius of the bend is $R=30$ nm [see Fig. \ref{1sc}(a)], unless stated otherwise.

The atomic spin-orbit interaction is due to the electric field of the nuclei. The spin-orbit interaction due to this field
mixes the $p_z$ orbitals with $p_x$, and $p_y$ orbitals of opposite spins.
The mixing  leads to hybridization of $p_z$ orbitals [labeled as $z_j$ for $j$-th ion -- see Fig. 1(a)]- forming the $\pi$ bonds with the $p_{x/y}$ (labeled as $x_j,y_j$)
-- forming the $\sigma$ bonds between the nearest neighbor atoms \cite{cnt2,ando,delvalle,martino}.
For $|\uparrow_{z_j}),\,|\downarrow_{z_j})$ standing for the spin-orbitals along the $z_j$ orbital of $j$th ion
the first-order corrections to the spin-orbital $|z_{j}\uparrow_{z_j}\rangle$
due to the atomic spin-orbit interaction stemming from the electric field of the nuclei have the form\cite{cnt2,ando,delvalle}:
\begin{equation}
|z_{j}\uparrow_{z_j}\rangle\approx|z_{j}\uparrow_{z_j})-\delta|x_{j}+iy_{j}\downarrow_{z_j}) \label{m1}
\end{equation}
and
\begin{equation}
|z_{j}\downarrow_{z_j}\rangle\approx|z_{j}\downarrow_{z_j})+\delta|x_{j}-iy_{j}\uparrow_{z_j}), \label{m2}
\end{equation} where $\delta$ parameter describes the strength of the SO coupling.
For the flat graphene the $z$ and $x/y$ orbitals for the nearest neighbor atoms are orthogonal, but
the orthogonality is lifted when the graphene plane is folded into the CNT \cite{ando,martino,schulz,izumida,jeong,chicoprl,chicoprb},
which introduces spin-orbit coupling effects to the $\pi$ band \cite{cnt2}.

The second-order tight-binding Hamiltonian theories taking into account the $2s,2p_x,2p_y,$ and $2p_z$ orbitals
are available \cite{izumida,klino1,chicoprl,chicoprb,jeong}. The contribution of the entire basis introduces combined effects
including the electron-hole asymmetry in the spectrum \cite{jeong,chicoprl,chicoprb} and spin-splitting of the bands for chiral CNT's \cite{izumida,chicoprb}.
In this paper we restrict the study to zigzag (achiral) CNT and consider the conduction band states only.
We aim at estimation of the spin-orbit coupling effects due to the bend of the entire CNT. Since
the effects of the bend should be expected much smaller than the ones due to folding of the graphene to the CNT we restrict
the modeling to the lowest-order corrections only \cite{ando,delvalle}.
Namely, we consider only the $p_z$ electrons, and the effects of mixing of type given by Eqs. (\ref{m1}) and (\ref{m2}).

We look for an effective \cite{ando,delvalle} tight-binding Hamiltonian of form
\begin{multline}
H=\sum_{\{i,j,\sigma,\sigma'\}}(c_{i\sigma}^\dagger \cdot t_{ij}^{\sigma\sigma'} \cdot  c_{j\sigma'}+h.c.)\label{eqh}\\
+\sum_{i,\sigma,\sigma'}c_{i\sigma}^\dagger  \cdot  \left(W({\bf r}_i)+\frac{g\mu_b}{2}  \boldsymbol{ \sigma}\cdot {\bf B}\right) \cdot c_{i\sigma'}.
\end{multline}
The first summation runs over $p_z$ spin-orbitals of nearest neighbor atoms, $c_{i\sigma}^\dagger$ $(c_{i\sigma})$ creates (annihilates) the electron at ion $i$ with spin $\sigma$ in $z$ direction, and  $t_{ij}^{\sigma\sigma'}$ is the spin-dependent hopping parameter. The second summation in Eq. (\ref{eqh}) accounts
for the external potential and the Zeeman interaction with the Land\`e factor $g=2$. $\boldsymbol{\sigma}$ stands for the vector of Pauli matrices, and {\bf B} for the magnetic field vector.
The magnetic field is necessary for the EDSR transitions to be observed. We assume that {\bf B} is applied in the $z$ direction [see Fig. 1(a)].

The double quantum dot (DQD) potential induced
by e.g. external gates is modeled as a sum of two Gaussians $V_{DQD}(z)=-V\left[\exp((z-s)^2/d^2)+\exp((z+s)^2/d^2)\right]$, where the distance between
the centers of the dots is $2s=10$ nm, and the width of a single QD is $2d=4.4$ nm. Note, that the DQD potential - depends on the global $z$ direction [see Fig. \ref{1sc}(a)]
and not on a coordinate along the length of the bent CNT.
Thus, the bend lowers the angular symmetry of the eigenstates with respect to the axis of the tube, which as discussed in Ref. \cite{eosprb} allows for the spin-flip transitions in
presence of the SO coupling.
The external potential defined within the CNT in the absence of the ac field is then $W=V_{DQD}(z)+e F_{bias}z$, where the second term is responsible for the bias field applied to the system.

The previous work \cite{eosprb} used the hopping parameters for a straight CNT, while the present paper deals with effects of the bend to the parameters including the SO interaction.
To include the effect of the bend into SO coupling we need to modify the hopping parameters. For necessary transformations we use four different coordinate systems (see Fig.  \ref{1sc}(a-c)): ($xyz$) - global
coordinate system, ($x_{j}y_{j}z_{j}$) - local coordinate system with $z_j$ aligned with $p_z$ spin-orbital at $j$-th ion and $y_j$ parallel to local nanotube axis, ($x_{\phi}y_{\phi}z_{\phi}$) -
auxiliary local coordinate system with $x_{\phi}$ coinciding with the global $y$ axis and $y_{\phi}$ coinciding with $y_j$ axis,  ($x'_{j}y'_{j}z'_{j}$) - system equivalent to $x_{j}y_{j}z_{j}$ but rotated
by $\phi$ angle about the $x_{\phi}$ axis.

The procedure for derivation of the hopping parameters for the bent nanotube
is the following:\newline
{\it i)} We consider each of the cross section of the nanotube (pink line on the Fig.  \ref{1sc}(a)), separately
and start from the eigenspinors $|\uparrow_{z_j}),\,|\downarrow_{z_j})$  defined along the $z_j$ orbital -- as in Ref. \cite{ando} [see Fig. \ref{1sc}(b)].\newline
{\it ii)} We rotate these eigenspinors by the $\phi$ angle about the axis $\boldsymbol{x_{\phi}}=(\cos\theta_{j},0,\sin\theta_{j})$
(see Fig. \ref{1sc}(b-c)). The rotation matrix has the form:
\begin{eqnarray}
{\bf A}=\cos\frac{\phi_{j}}{2}\boldsymbol{I}-i\sin\frac{\phi_{j}}{2}\boldsymbol{x_{\phi}}\cdot\boldsymbol{\sigma_{j}}=\nonumber \\ \left[\begin{array}{cc}
\cos\frac{\phi_{j}}{2}-i\sin\frac{\phi_{j}}{2}\sin\theta_{j} & -i\sin\frac{\phi_{j}}{2}\cos\theta_{j}\\
-i\sin\frac{\phi_{j}}{2}\cos\theta_{j} & \cos\frac{\phi_{j}}{2}+i\sin\frac{\phi_{j}}{2}\sin\theta_{j}
\end{array}\right]
\end{eqnarray}
with $\boldsymbol{\sigma_{j}}$ -- the Pauli matrices in the local coordinate
system $x_{j}y_{j}z_{j}$. We get new eigenspinors in the rotated
system $x'_{j}y'_{j}z'_{j}$ with axis $y'_{j}$ coinciding with the global $z$ direction:
\begin{equation}
|\uparrow_{z'_j})=(\cos\frac{\phi_{j}}{2}-i\sin\frac{\phi_{j}}{2}\sin\theta_{j})|\uparrow_{z_j})-i\sin\frac{\phi_{j}}{2}\cos\theta_{j}|\downarrow_{z_j})
\end{equation}
\begin{equation}
|\downarrow_{z'_j})=-i\sin\frac{\phi_{j}}{2}\cos\theta_{j}|\uparrow_{z_j})+(\cos\frac{\phi_{j}}{2}+i\sin\frac{\phi_{j}}{2}\sin\theta_{j})|\downarrow_{z_j})
\end{equation}
{\it iii)} We transform the eigenspinors $|\uparrow_{z'_j})$, $|\downarrow_{z'_j})$ along the $z'_j$
direction in the $x'_{j}y'_{j}z'_{j}$ coordinate system into eigenspinors in $z$ direction in the $xyz$ coordinate system \cite{delvalle}
$|\uparrow_{z}),\,|\downarrow_{z})$ [see Fig. \ref{1sc}(a,c)]:
\begin{eqnarray}
|\uparrow_{z}\rangle&=& \frac{e^{i\theta_{j}/2}}{\sqrt{2}}\{|\uparrow_{z'_j})+i|\downarrow_{z'_j})\} \nonumber \\
&=&\frac{1}{\sqrt{2}}\{(\cos\frac{\phi_{j}}{2}e^{i\theta_{j}/2}+\sin\frac{\phi_{j}}{2}e^{-i\theta_{j}/2})|\uparrow_{z_j})\nonumber \\ &+&i(\cos\frac{\phi_{j}}{2}e^{i\theta_{j}/2}-\sin\frac{\phi_{j}}{2}e^{-i\theta_{j}/2})|\downarrow_{z_j})\},
\end{eqnarray}
and
\begin{eqnarray}
|\downarrow_{z}\rangle&=&\frac{e^{-i\theta_{j}/2}}{\sqrt{2}}\{|\uparrow_{z'_j})-i|\downarrow_{z'_j})\}\nonumber \\&=&
\frac{1}{\sqrt{2}}\{(\cos\frac{\phi_{j}}{2}e^{-i\theta_{j}/2}-\sin\frac{\phi_{j}}{2}e^{i\theta_{j}/2})|\uparrow_{z_j})\nonumber \\&-&i(\cos\frac{\phi_{j}}{2}e^{-i\theta_{j}/2}+\sin\frac{\phi_{j}}{2}e^{i\theta_{j}/2})|\downarrow_{z_j})\}.
\end{eqnarray}
{\it iv)} Using transformations {\it iii)} we convert formulae (\ref{m1}) and (\ref{m2}) to
the $|z_{j}\uparrow_{z}\rangle$ basis:
\begin{eqnarray}
|z_{j}\uparrow_{z}\rangle&\approx&|z_{j}\uparrow_{z})+i\delta\cos\phi_{j}|x_{j}\uparrow_{z})-i\delta\sin\phi_{j}|x_{j}\downarrow_{z})\nonumber \\
&+&i\delta\sin\phi_{j}\sin\theta_{j}|y_{j}\uparrow_{z})\\ &+&\delta(\cos^{2}\frac{\phi_{j}}{2}e^{i\theta_{j}}+\sin^{2}\frac{\phi_{j}}{2}e^{-i\theta_{j}})|y_{j}\downarrow_{z}) \nonumber ,
\end{eqnarray}
and
\begin{eqnarray}
|z_{j}\downarrow_{z}\rangle&\approx&|z_{j}\downarrow_{z})-i\delta\sin\phi_{j}|x_{j}\uparrow_{z})-i\delta\cos\phi_{j}|x_{j}\downarrow_{z}) \nonumber \\
&-&\delta(\cos^{2}\frac{\phi_{j}}{2}e^{-i\theta_{j}}+\sin^{2}\frac{\phi_{j}}{2}e^{i\theta_{j}})|y_{j}\uparrow_{z})\\ &-&i\delta\sin\phi_{j}\sin\theta_{j}|y_{j}\downarrow_{z})\nonumber
\end{eqnarray}
{\it v)} Finally, using {\it iv)} and discarding higher-order terms in $\delta$ we obtain the hopping parameters to be used in Hamiltonian (1):
\begin{eqnarray}
t_{ij}^{\uparrow\uparrow}&=&\langle z_{i}\uparrow_{z}|H|z_{j}\uparrow_{z}\rangle\nonumber\\ &=&(z_{i}|H|z_{j})\nonumber +i\delta\cos\phi_{j}(z_{i}|H|x_{j})-i\delta\cos\phi_{i}(x_{i}|H|z_{j}) \nonumber \\
&+&i\delta\sin\phi_{j}\sin\theta_{j}(z_{i}|H|y_{j})  \nonumber \\
&-&i\delta\sin\phi_{i}\sin\theta_{i}(y_{i}|H|z_{j}),
\end{eqnarray}
\begin{eqnarray}
t_{ij}^{\downarrow\downarrow}&=&\langle z_{i}\downarrow_{z}|H|z_{j}\downarrow_{z}\rangle\nonumber \\
&=&(z_{i}|H|z_{j}) -i\delta\cos\phi_{j}(z_{i}|H|x_{j})+i\delta\cos\phi_{i}(x_{i}|H|z_{j}) \nonumber \\
&-&i\delta\sin\phi_{j}\sin\theta_{j}(z_{i}|H|y_{j})  \nonumber \\
&+&i\delta\sin\phi_{i}\sin\theta_{i}(y_{i}|H|z_{j}),
\end{eqnarray}
\begin{eqnarray}
t_{ij}^{\uparrow\downarrow}&=&\langle z_{i}\uparrow_{z}|H|z_{j}\downarrow_{z}\rangle\nonumber \\
&=&-i\delta\sin\phi_{j}(z_{i}|H|x_{j})+i\delta\sin\phi_{i}(x_{i}|H|z_{j})\nonumber \\
&-&\delta(\sin^{2}\frac{\phi_{j}}{2}e^{i\theta_{j}}+\cos^{2}\frac{\phi_{j}}{2}e^{-i\theta_{j}})(z_{i}|H|y_{j})\nonumber \\&+&\delta(\sin^{2}\frac{\phi_{i}}{2}e^{i\theta_{i}}+\cos^{2}\frac{\phi_{i}}{2}e^{-i\theta_{i}})(y_{i}|H|z_{j}),
\end{eqnarray}
and
\begin{eqnarray}
t_{ij}^{\downarrow\uparrow} &=& \langle z_{i}\downarrow_{z}|H|z_{j}\uparrow_{z}\rangle\nonumber \\
&=&-i\delta\sin\phi_{j}(z_{i}|H|x_{j})+i\delta\sin\phi_{i}(x_{i}|H|z_{j})\nonumber\\
&+&\delta(\sin^{2}\frac{\phi_{j}}{2}e^{-i\theta_{j}}+\cos^{2}\frac{\phi_{j}}{2}e^{i\theta_{j}})(z_{i}|H|y_{j})\nonumber \\&-&\delta(\sin^{2}\frac{\phi_{i}}{2}e^{-i\theta_{i}}+\cos^{2}\frac{\phi_{i}}{2}e^{i\theta_{i}})(y_{i}|H|z_{j}).
\end{eqnarray}

Following Ref. \cite{ando} the matrix elements for the neighbor $p_z$ orbitals to be used in the hopping parameters above read
\begin{eqnarray}
(\alpha_{i}|H|\alpha_{j})&=&V_{pp}^{\pi}{\bf n}(\alpha_{i})\cdot{\bf n}(\alpha_{j})\\
&+&(V_{pp}^{\sigma}-V_{pp}^{\pi})\frac{({\bf n}(\alpha_{i})\cdot{\bf R}_{ji})({\bf n}(\alpha_{j})\cdot{\bf R}_{ji})}{{\bf |R}_{ji}|^{2}},\nonumber
\end{eqnarray}
where $\alpha=x,\, y$ or $z$, $\alpha_{i}$ is orbital localized
at site ${\bf R}_{i}$ ($i$-th ion), ${\bf n}(\alpha_{i})$ is a unit
vector in the direction of orbital $\alpha_{i}$  [see Fig. \ref{1sc}(b)].

For a straight CNT, $\phi_j=0$ for all $j$, the formulae reduce to the form given by Ref. \cite{delvalle},
for which non-zero spin-flipping hopping parameters appear only along the circumference of the nanotube ($\theta_i\neq \theta_j$), where the
$p_z$ orbitals are no longer strictly parallel. The form of spin-orbit coupling derived above
accounts for the deflection of $p_z$ orbitals along the length of the bent CNT, and the spin-flipping hopping parameters
appear also for neighbor orbitals with same $\theta$ but varied $\phi$.
In the numerical results below we use the spin-orbit coupling parameter $\delta=0.003$ after Refs. \cite{ando,delvalle}.
and the tight-binding parameters $V_{pp}^{\pi}=-2.66$ eV and $V_{pp}^\sigma=6.38$ of Ref. \cite{tomanek}

The magnetic field induces the spin Zeeman effect in Hamiltonian (1).
The orbital effects of the magnetic field are introduced by Peierls phase to the hopping terms $t_{ij}(B)= t_{ij}(0)\exp(i\frac{2\pi}{\Phi_0} \int_{{\bf r}_i}^{{\bf r}_j}{\bf A}\cdot {\bf dl}),$
where ${\bf B}=\nabla \times {\bf A}$, and $\Phi_0=h/e$ is the flux quantum.

Below we discuss the contribution of the SO coupling due to the bend.
The reference results are obtained
for the hopping parameters as for the straight CNT \cite{delvalle} and
the bend of the tube enters only the potential energy and the vector potential through the spatial coordinates of the ions.

\section{Stationary states}
The calculated energy spectrum of the CNT is plotted in Figure \ref{1sc}(d) for $F_{bias}=10$ kV/cm. With the black points we plotted the energy levels of the states which are localized
outside the double quantum dot potential as functions of the depth of the Gaussian cavities. The two horizontal black lines correspond to
energy levels localized at the zigzag edges of the tube.
The color lines indicate the energy levels localized within the DQD, the red (green) ones with the majority
of the charge localized in the left $z<0$ (right $z>0$) dot.  Below we assume $V=0.55$ eV, for which one red and one green energy levels -- each nearly four-fold degenerate
with respect to the valley and the spin, appear below the neutrality point energy (zero). We will consider a single-electron localized in DQD inside a CNT which is
otherwise neutral.

In Figure \ref{1sc}(f) we plotted the energy levels as a function of the bias field.
Each of the plotted energy levels is two-fold degenerate. At $F_{bias}=0$ the splitting between the green and red energy levels is due to the tunnel coupling between the dots
and formation of bonding and antibonding-orbitals. Each pair of the energy levels is split by the spin-orbit coupling energy. As the bias is applied the lower (upper) quadruple
of energy levels becomes localized in the left (right) dot. The electron localization of the energy levels is given in Figure \ref{1sc}(g-h) for the bias fields $F_{bias}=2$ kV/cm and 10 kV/cm.

The magnetic field splits the degeneracy of the energy levels -- see the energy spectrum of Fig. \ref{1sc}(e).
We label the states by the valley \cite{bulaev} the majority spin component $\uparrow\downarrow$
and the dot left or right ($l,r$) in which most of the charge is localized.
Below we consider transitions from the ground-state $K'\uparrow$ to the excited states at $B=5$T.
The present calculation accounts for both the valleys, but since the modelled CNT is defect-free no intervalley
transitions are observed.
We discuss transitions from  $K'$ ground-state to the final states of the $K'$ valley.

\begin{figure}[h]
\begin{tabular}{l}
\includegraphics[width=0.65\textwidth]{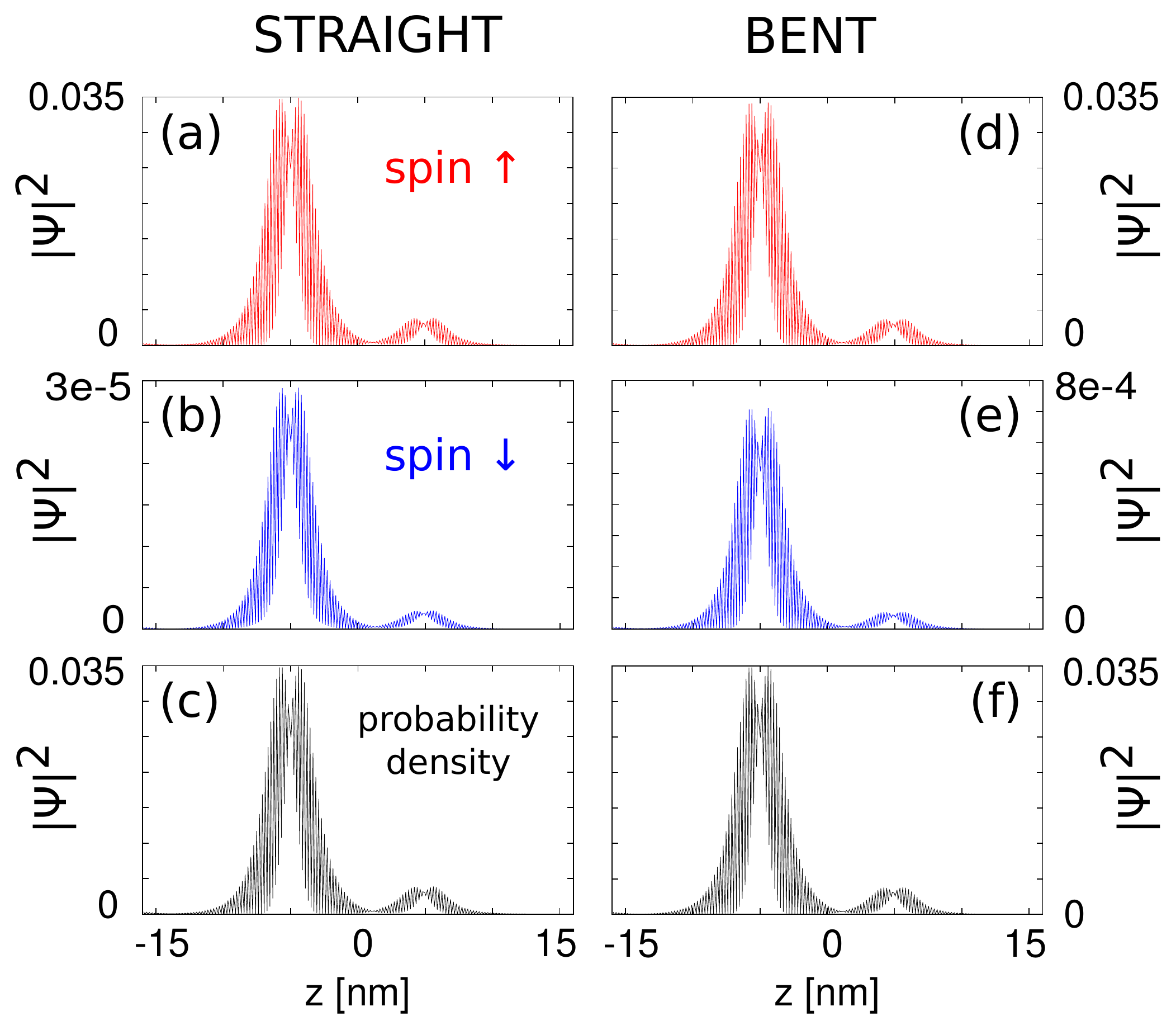}
\end{tabular}
\caption{\footnotesize \linespread{0.5} Charge (c,f) and spin (a-e) densities for the two-electron ground state at $F_{bias}=10$ kV/cm and $B=5$T.
Results correspond to the bend nanotube. In (a-c) SO coupling (hopping parameters) was adopted for a straight CNT.
The SO that accounts for the bend was used in (d-f). Arbitrary units for the probability density are used. } \label{gs}
\end{figure}

The spin-orbit coupling \cite{ando} which stems from the finite radius $r$ of the nanotube results in formation of the spin-valley doublets split by the spin-orbit interaction energy $\Delta_{SO}$.
The energy effect of the spin-orbit coupling resulting from a  finite radius of the bend $R$ is too small to be plotted in Fig. \ref{1sc}(e) and for $B=5$T it is of the order of 0.01 meV at most,
depending on the energy level.
A relative effect on the average spin of the states is larger. For instance $\langle s_z\rangle$ for the ground state at 5T is reduced by the spin-orbit coupling due to the bend from $0.4991\hbar$ to $0.4805\hbar$.
Figure \ref{gs} compares the ground-state spin and charge density obtained for SO coupling for a straight CNT with the ones including the effect of the bend.
The charge and spin distribution are similar in shape, but the minority spin density is increased by about 25 times by the SO due to the bend.

\begin{figure*}[htbp]
\begin{tabular}{lr}
\includegraphics[width=0.47\textwidth]{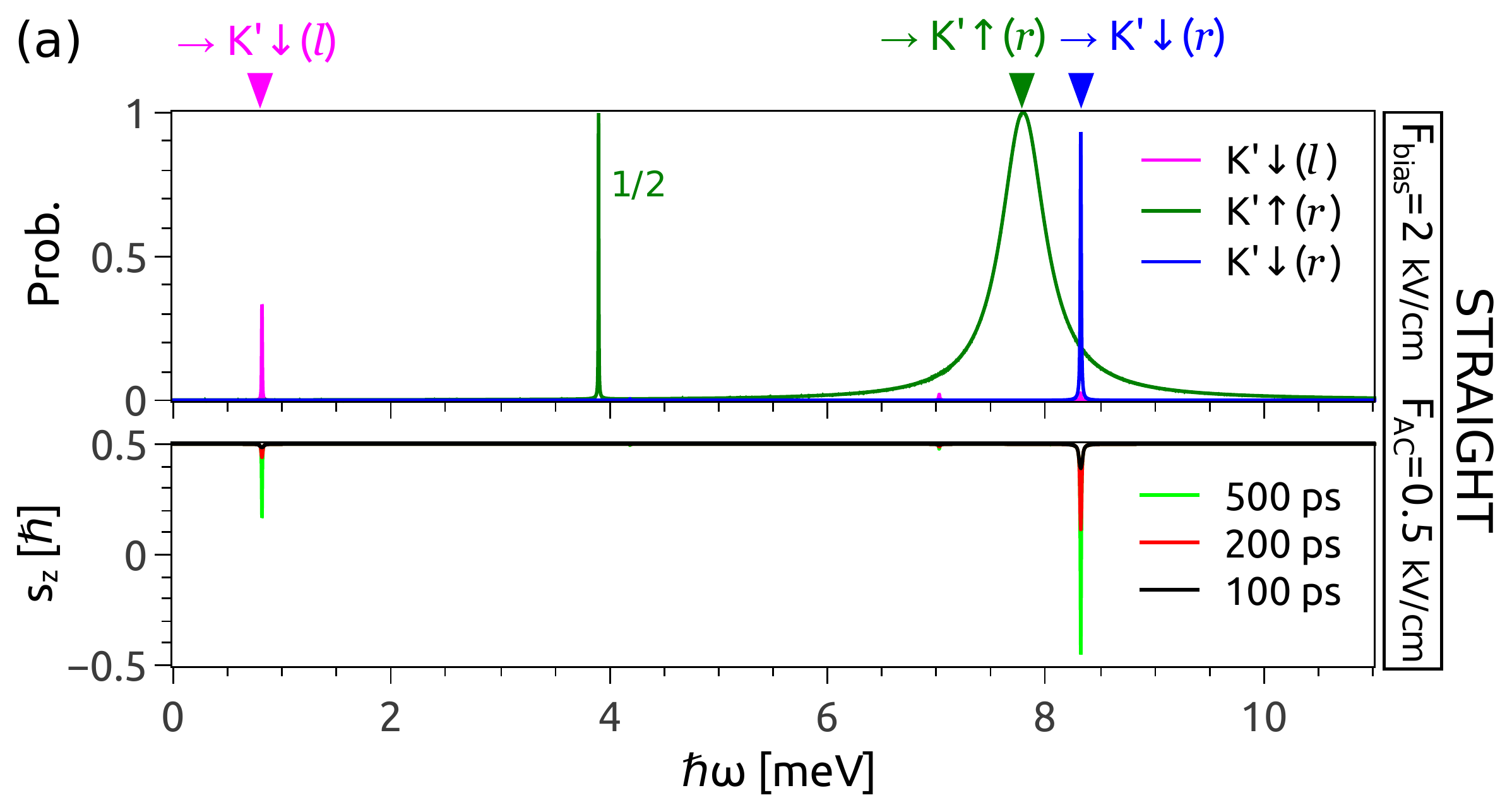} &  \includegraphics[width=0.47\textwidth]{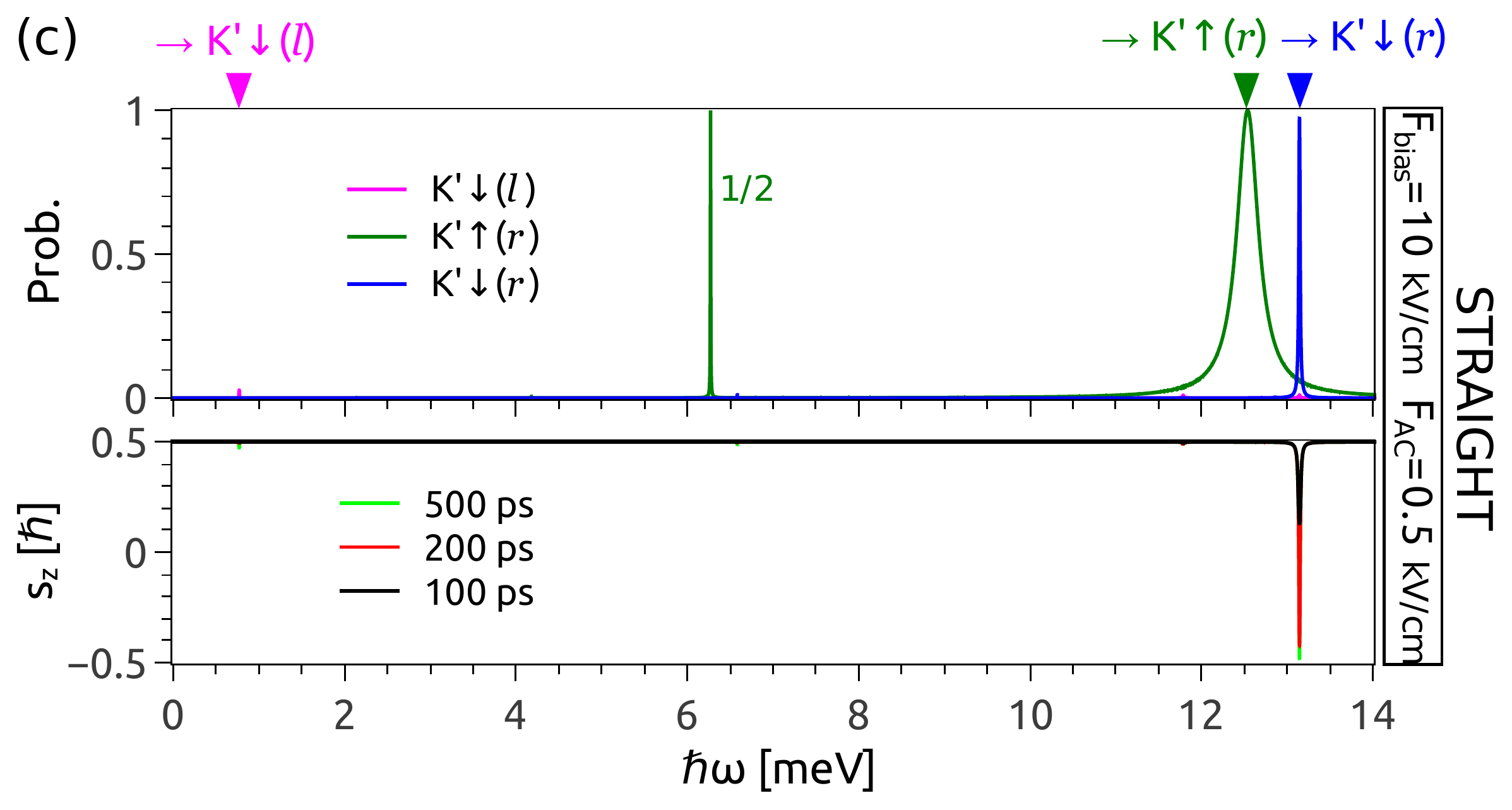}\\
\includegraphics[width=0.47\textwidth]{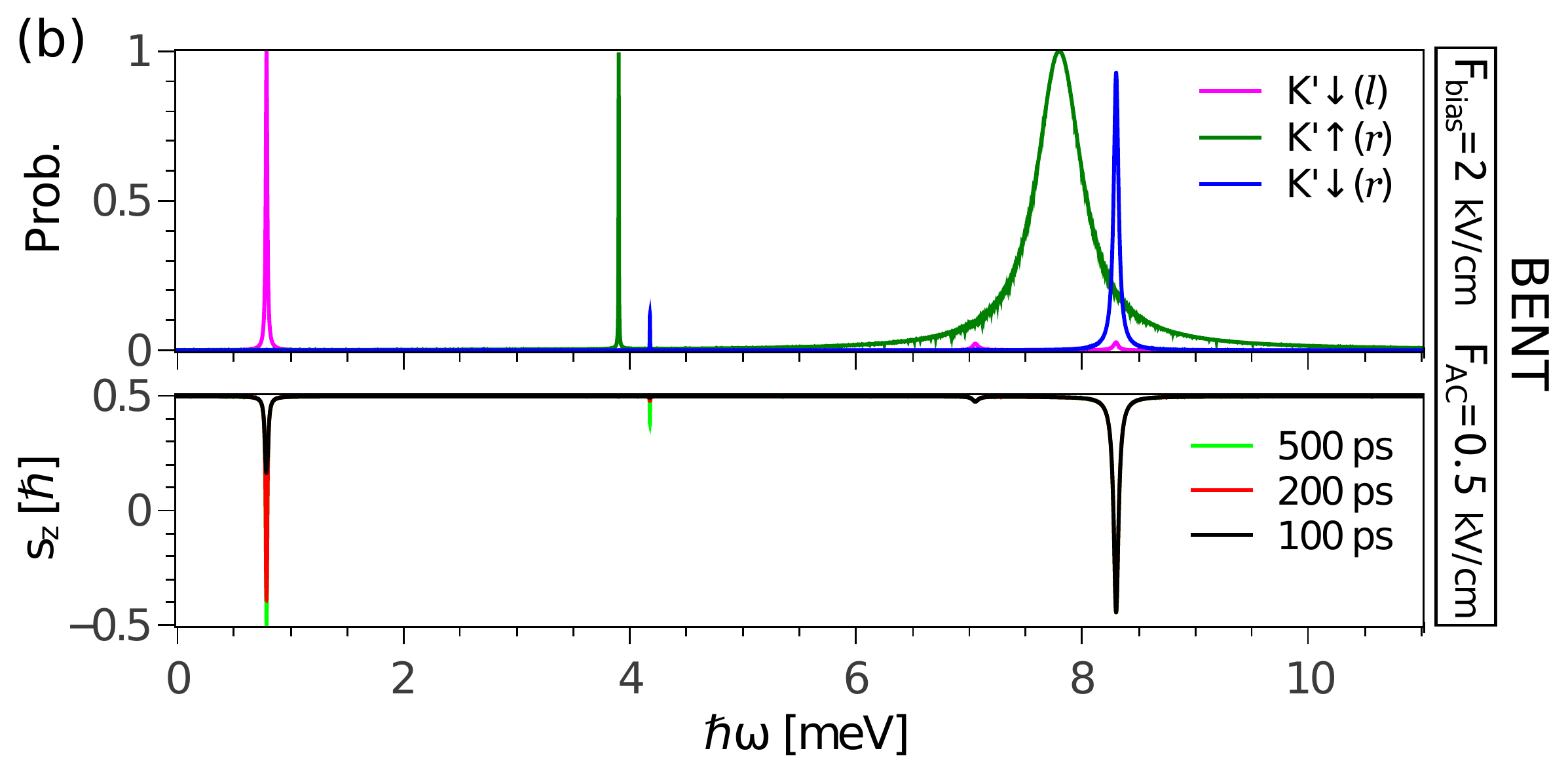} &  \includegraphics[width=0.47\textwidth]{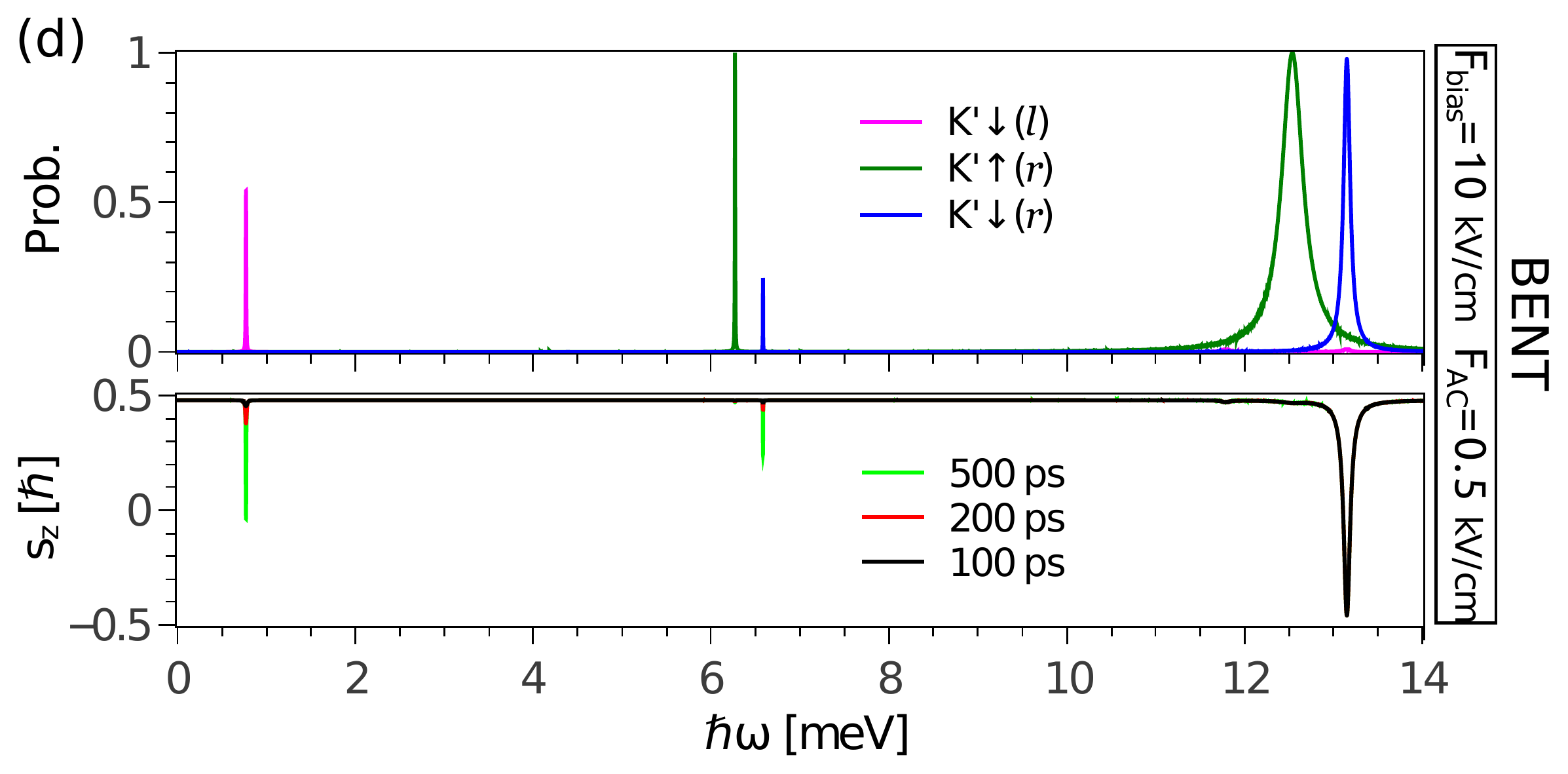}\\
\\
\\
\includegraphics[width=0.47\textwidth]{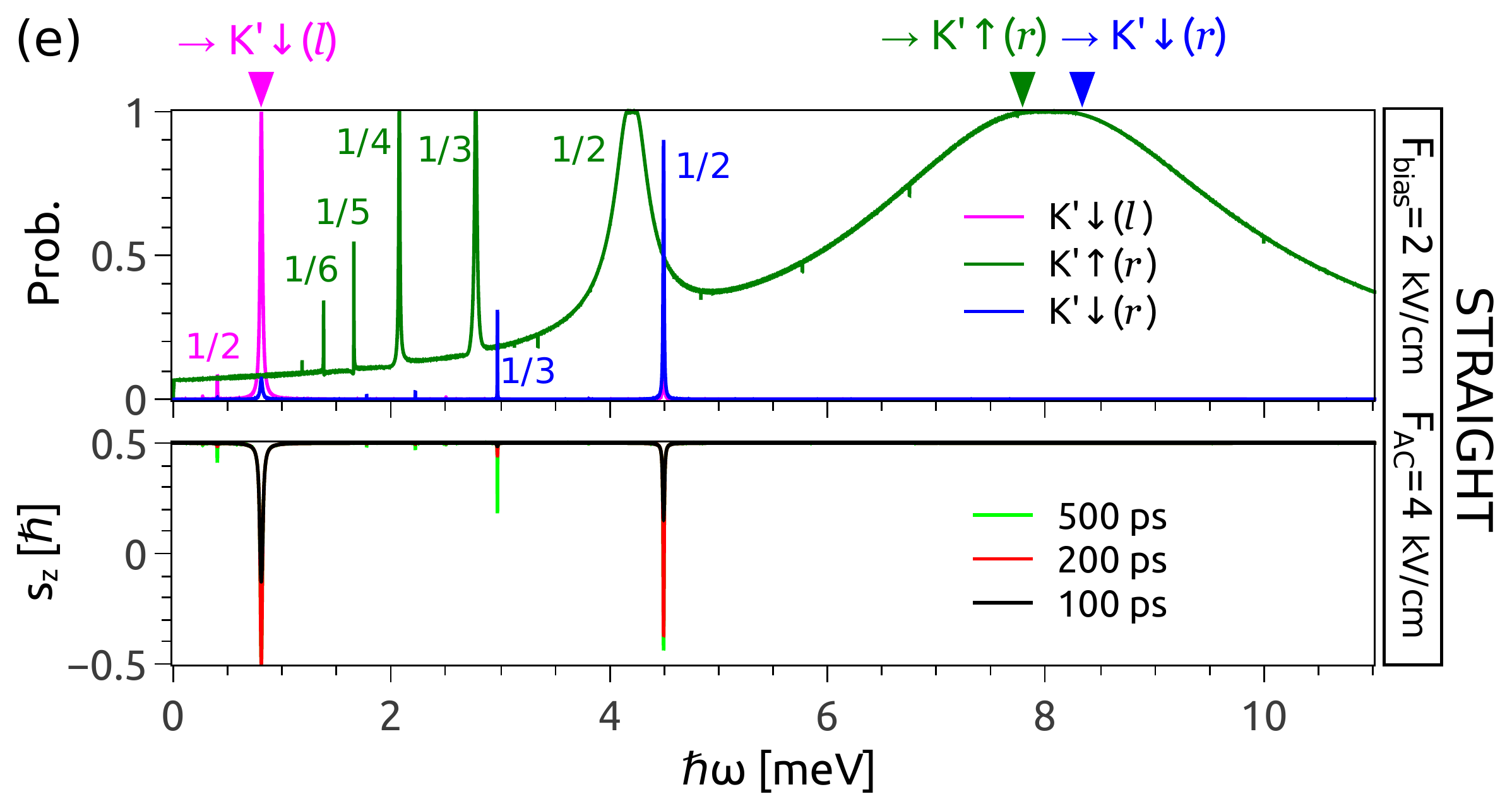} & \includegraphics[width=0.47\textwidth]{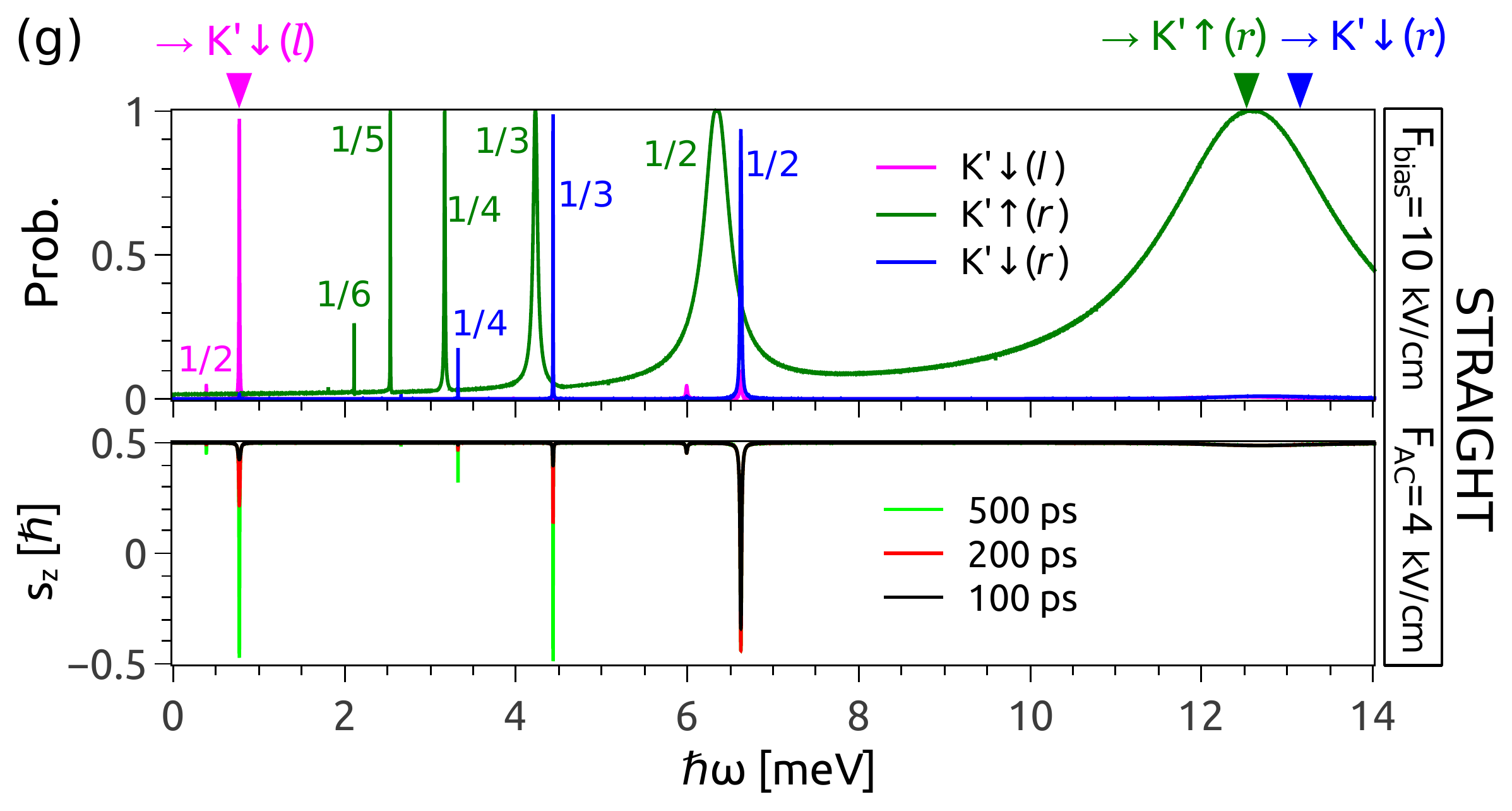}\\
\includegraphics[width=0.47\textwidth]{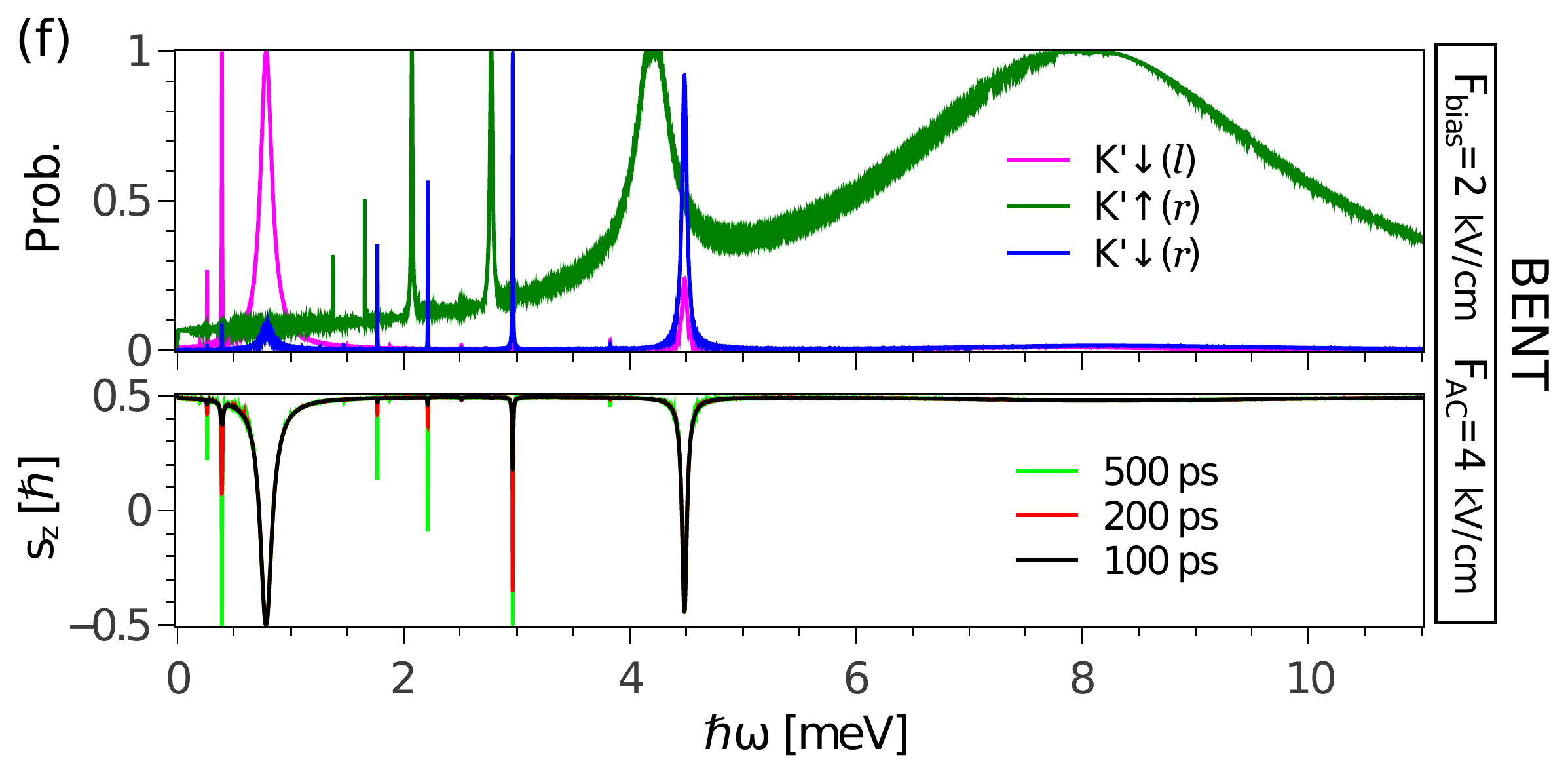} & \includegraphics[width=0.47\textwidth]{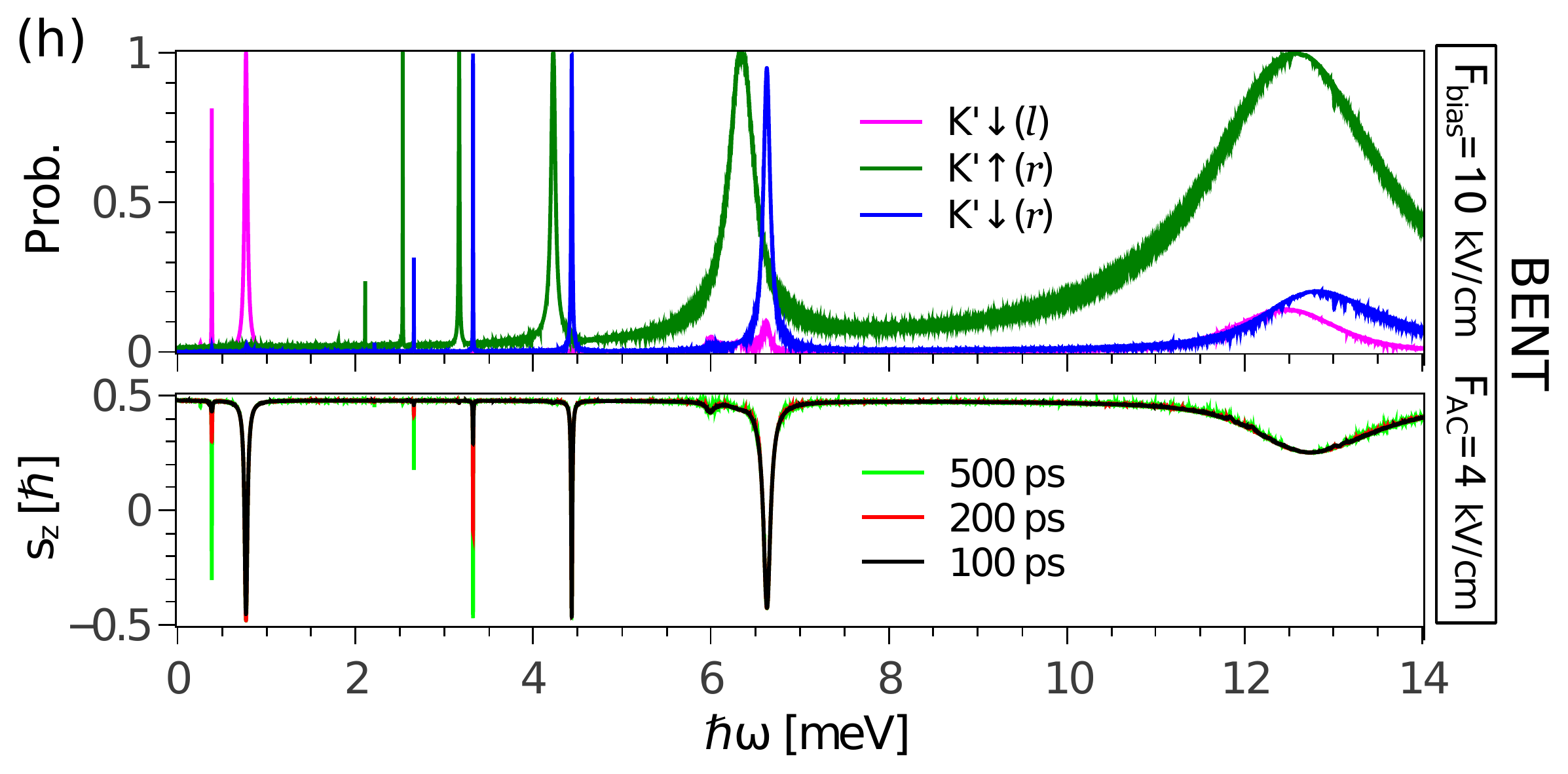}\\
\end{tabular}
\vspace{0.4cm}
\caption{\footnotesize \linespread{0.5} Upper panels of (a-h): maximal occupation probability for the three $K'$ excited energy levels reached during 500 ps time evolution
from the $K'\uparrow(l)$ ground state in the initial condition. Pink/green/blue lines correspond to the transitions to $K'\downarrow(l)/K'\uparrow(r)/K'\downarrow(r)$ states, respectively.
The triangles on top of the upper axis indicate the energy difference between the ground-state and the excited $K'$ states
as calculated from the energy spectrum. The lower panels of (a-h): minimal spin obtained during the evolution. Light green lines indicate minimal spin for 500 ps simulations (corresponding to results from
upper panels), red/black lines show intermediate results for 200 ps/100 ps simulations.
The applied bias is $F_{bias}=2$ kV/cm (a-b,e-f) and $F_{bias}=10$ kV/cm (c-d,g-h).
The amplitude of the ac field is $F_{AC}=0.5$ kV/cm in (a-d) and $F_{AC}=4$ kV/cm in (e-h).
Plots (a,c,e,g) correspond to the results obtained with the hopping parameters neglecting the bend and (b,d,f,h) with hopping parameters including the bend.
The plotted scans were calculated with a $\hbar\omega$ spacing of 0.1$\mu$eV.
} \label{1scan}
\end{figure*}

\section{Spin and charge transitions}

We investigate the spin and charge dynamics of the system when an external ac electric field is applied along the $z$ direction,
$V_{AC}(t)=eF_{AC}z\sin(\omega t)$. For that purpose we solve the Schr\"odinger equation $i\hbar\frac{d\Psi}{dt}=H'\Psi$ for Hamiltonian $H'(t)=H+V_{AC}(t)$ in the basis of $H$ eigenstates,
\begin{equation}
\Psi({\bf r},\sigma,t)=\sum_{n=1}^Nc_{n}(t) \Psi_{n}({\bf r},\sigma)e^{-\frac{iE_{n}t}{\hbar}}, \label{basa}
\end{equation}
where $H\Psi_n=E_n\Psi_n$. For the wave function (\ref{basa}) introduced to the Schr\"odinger equation, after application of the Galerkin projection, the time
evolution is given by a system of ordinary differential equations
\begin{equation}
i\hbar \dot{c}_{k}(t)=\sum_{n=1}^Nc_{n}(t)eF_{AC}\sin(\omega t)\langle \Psi_{k}|z|\Psi_{n}\rangle e^{-\frac{i(E_{n}-E_{k})t}{\hbar}},
\end{equation}
for $k=1,2,\dots N$ that we solve with the Crank-Nicolson time-stepping method.
We consider that all the energy levels below the ground-state level localized within the DQD, with energies $E<-300$ meV for $V=0.55$ eV [see Fig. 1(d)] are
fully occupied. For the time evolution we take $N=72$ lowest-energy spin-orbitals to the basis (\ref{basa}) starting from the DQD ground-state.
The results can be considered as exact solutions of the time dependent Schr\"odinger equation since
introduction of a larger number of eigenstates to the basis does not alter the results in a detectable manner.

\subsection{Spin-conserving transitions}
The ac electric field drives the transitions between the four $K'$ valley states. Figure \ref{1scan}
 shows [upper panels of Fig. \ref{1scan}(a-h)] the maximal projection ($|\langle \Psi(t)|\Psi_n \rangle|^2$)  of the evolving quantum state starting from the ground $K'\uparrow(l)$ state on the three excited $K'$ states
for simulation that covers 500 ps.   The green line in the upper panels of Fig. \ref{1scan} corresponds to the spin-conserving transition $K'\uparrow(l)\rightarrow K'\uparrow(r)$ with the electron passing
from the left to the right quantum dot. The spin-conserving charge transition is very fast. At the bias voltage $F_{bias}=2$ kV/cm  for the resonant energy $\hbar\omega$ equal to the energy difference
between the $K'\uparrow(l)$ and $K'\uparrow(r)$ energy levels the charge transition from the left to the right dot appears within about 8 ps ($F_{AC}=0.5$ kV/cm).
For $F_{AC}$ increased to 4 kV/cm the corresponding transition time is about 1 ps only.
The charge transition between the dots slows down when the bias increases the detuning of the single-dot energy levels and reduces the overlap of the spatial wave functions of the initial
and final states. For $F_{bias}=10$ kV/cm the charge transition times are $\simeq 14$ ps
and $\simeq 2$ ps for $F_{AC}=0.5$ kV/cm and $F_{AC}=4$ kV/cm, respectively.
The lower rate of the charge transition is consistent with the dipole matrix elements of Table I,
and is found also for the CNT with SO including the effect of the bend.
Note, that the amplitude of the ac field has a pronounced influence on the width of transition $K'\uparrow(l) \rightarrow K'\uparrow(r)$ [cf. Fig. \ref{1scan}(a,e) and Fig. \ref{1scan}(c,g)].
For a small amplitude of the alternate electric fields $F_{AC}=0.5$ kV/cm we observe [upper panel of Fig. \ref{1scan}(a,c)] a narrow resonant transition at half the
energy spacing between the ground state $K'\uparrow(l)$ and the excited state $K'\uparrow(r)$ (see the peak marked by 1/2 in the Figures).
This is the fractional resonance, which is a counterpart of two-photon transitions that are observed for atoms and molecules in strong laser fields \cite{lewenstein}.
When the amplitude of the ac field is increased, one observes a series of higher-order resonances with the resonant amplitude
up to 6th harmonics [upper panel of Fig. \ref{1scan}(e,g)].

\begin{table}[htbp]
\begin{tabular}{|c|c|c|c|c|}
\hline
$R$ [nm] &$F_{bias}$& $K'\downarrow(l)$ & $K'\uparrow(r)$ & $K'\downarrow(r)$  \\ \hline
\multirow{2}{*}{15}&2 kV/cm &  0.068 (0.347) & 4.815 (4.666) & 0.180 (1.078) \\
& 10 kV/cm & 0.016  (0.074) & 2.852 (2.340) & 0.317 (1.565)\\ \hline
\multirow{2}{*}{30}&2 kV/cm &  0.032 (0.164) & 4.847 (4.820) & 0.090 (0.469) \\
& 10 kV/cm &  0.009 (0.043) & 3.010 (2.877) & 0.175 (0.859)\\ \hline
\multirow{2}{*}{120}&2 kV/cm & 0.008 (0.040) & 4.857 (4.856) & 0.023 (0.112) \\
& 10 kV/cm & 0.002 (0.011) & 3.059 (3.051) & 0.044 (0.220)\\ \hline
\end{tabular}
\caption{\footnotesize \linespread{0.5} Dipole matrix elements $|\langle K'\uparrow(l)|ez|\Psi \rangle |$ for the transition from the ground state to one of the three excited
states of the same valley, in units of $e\times$ nm for $B=5$ T. Results without (with) the inclusion of the SO coupling due to the bend of the nanotube
with radius $R$ are given outside (inside) parentheses.
}\end{table}

\begin{figure}[h]
\begin{tabular}{l}
\includegraphics[width=0.6\textwidth]{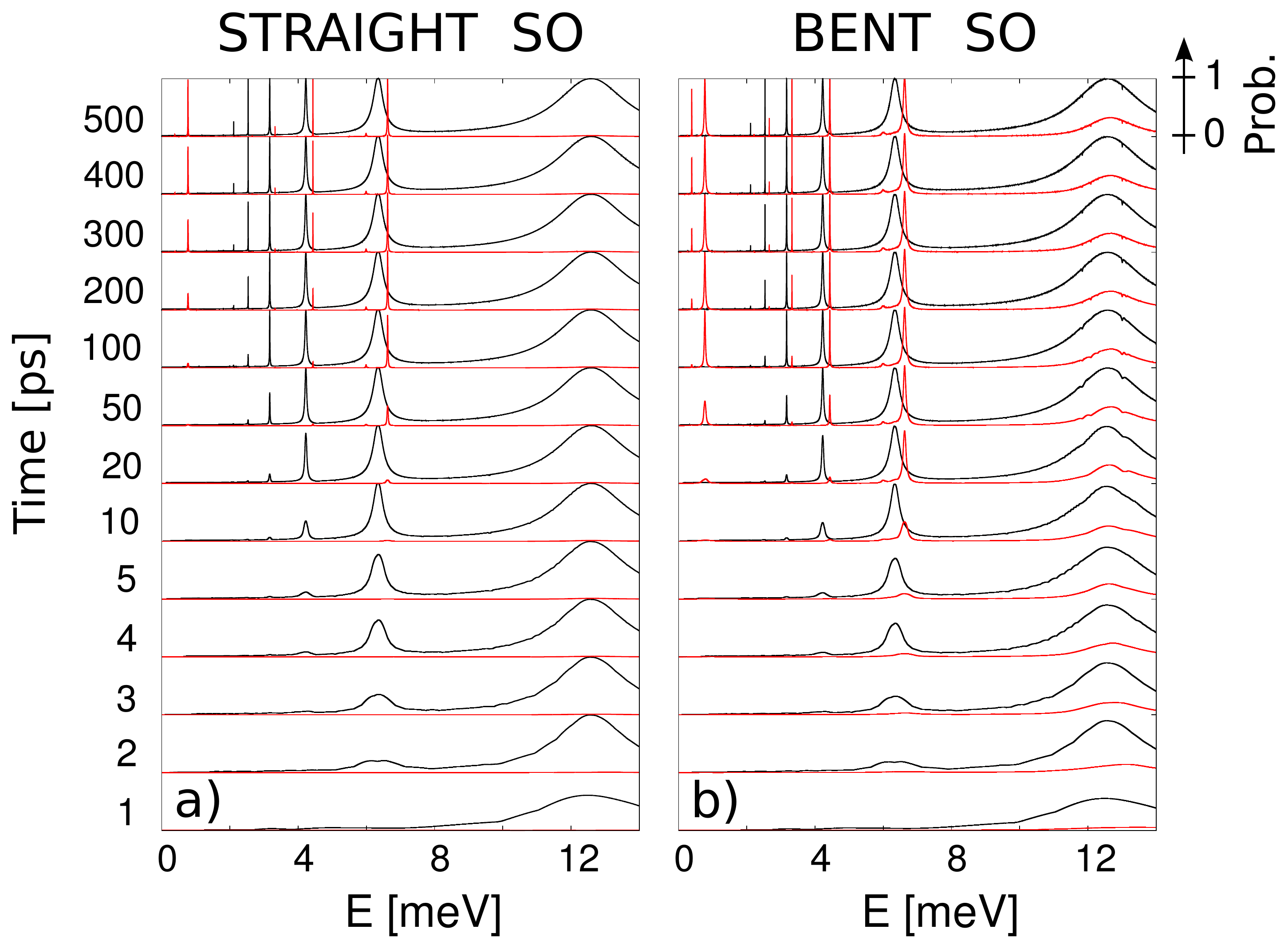}
\end{tabular}
\caption{\footnotesize \linespread{0.5} Spin conserving (black lines) and spin flipping (red lines) transition probabilities for a varied
duration of the AC field. In (a) the hopping parameters of a straight CNT are used. The bend of the CNT is included in (b). We use $F_{bias}=10$ kV/cm and $F_{AC}=4$ kV/cm as in Fig. 3(g,h). } \label{nowyfig}
\end{figure}

\subsection{Spin-flipping transitions}
Table I lists the dipole matrix elements for the transitions, which are inversely proportional to the transition times.
The oscillator strength for the transition with conserved spin $\rightarrow K'\uparrow(r)$ exceeds by  orders of magnitude
the ones for spin-flipping transitions.
 For a larger bias an increase of the matrix element for transition with spin-flip accompanied by interdot  charge hopping
 $K'\uparrow(l) \rightarrow K'\downarrow(r)$  is observed. An opposite tendency is found for the intradot spin flip $K'\uparrow(l) \rightarrow K'\downarrow(l)$.
In consistence with the data for the matrix elements given in Table I, the time-dependent calculations indicate that
for low amplitude of $F_{AC}=0.5$ kV/cm [Fig. \ref{1scan}(a,c)] the interdot spin flip is by far a more pronounced
transition than the spin flip with the electron staying in the left dot.
For SO coupling taken for the straight CNT the latter is weak at low bias [Fig. \ref{1scan}(a)] and nearly absent for the larger bias [Fig. \ref{1scan}(c)].

The fact that the spin-flip interdot transition occurs faster then the intra-dot spin inversion is quite counterintuitive,
but can be explained in simple terms based on the approximate symmetries of the wave functions. We find that the lower-energy states that are mostly localized
in the left dot have a bonding character,
i.e. {\it both} the majority $\uparrow$ and the minority $\downarrow$ components have a positive average value
of the z-parity operator $P_z$ defined as $P_zf(x,y,z)=f(x,y,-z)$.
On the other hand, the $K'\uparrow\downarrow(r)$ excited states have an antibonding character of both
the spin components with a negative average value for $P_z$. The transitions matrix elements (Table I) between pairs of states of opposite character (bonding-antibonding) in terms of formation
of artificial molecular orbitals are naturally larger than for the states of the same (bonding-bonding) character.
Therefore, the intradot transition $K'\uparrow(l)\rightarrow K'\downarrow(r)$ has a larger matrix element than the interdot one $K'\uparrow(l)\rightarrow K'\downarrow(l)$.

\begin{figure}[h]
\begin{tabular}{ll}
\includegraphics[width=0.5\textwidth]{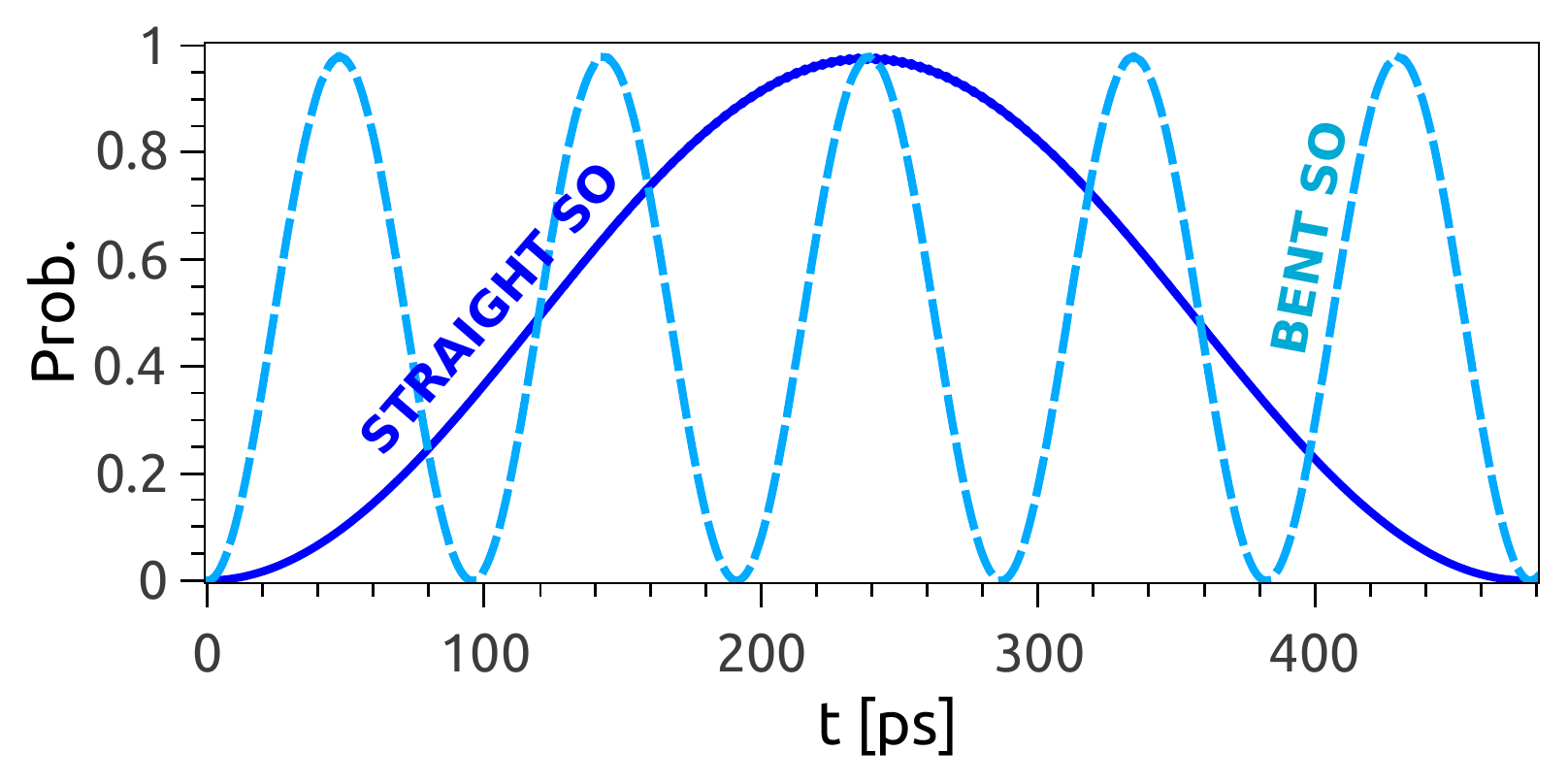}
\end{tabular}
\caption{\footnotesize \linespread{0.5} $K'\uparrow(l) \rightarrow K'\downarrow(r)$ transition: square of the absolute value of the wave function projection on the $K'\downarrow(r)$ state as a function of time for bias $F=10$ kV/cm and the driving amplitude $F_{AC}=0.5$ kV/cm.
The dashed (solid) curve corresponds to the results obtained with the hopping parameters including the bend (neglecting the bend).
The frequency was tuned to resonance -- see Fig.  \ref{1scan}(c-d) for each case.} \label{tflip}
\end{figure}

Let us now compare the results as obtained for SO interaction as set for a straight CNT [Fig. \ref{1scan}(a,c,e,g)],
and the ones for the SO theory accounting for the bend [Fig. 3(b,d,f,h)].
The spin-orbit interaction due to the bend has no significant influence on the rate or width of spin-conserving transitions.
The bend-related SO interaction shortens the direct transition rate to the $K'\downarrow$ state -- see the minimal $\langle s_z\rangle$
for 100, 200 and 500 ps simulation.
For $F_{bias}=10$ kV/cm  the transition to $K'\downarrow(r)$ state is displayed in a closer detail in Fig. \ref{tflip} as a function of time
for SO as obtained for a straight and bent CNT. The transition time is reduced 5 times by the contribution of the bend to the SO interaction,
in consistence with the results  for the matrix elements given in Table I.
In Fig. \ref{nowyfig} we plotted the induced transition probability without (a) and with (b) the bend-related contribution to SO interaction for parameters of Fig. 3(g,h) as a function of time. For identification of the separate lines -- see Fig. 3(g,h). The fully developed spin-flip and charge hopping line that is related to the half-resonant transition (see the thin red line for $E=6.3$ meV) requires 20 ps [Fig. \ref{nowyfig}(a)], while the theory neglecting the bend contribution to the SO interaction predicts transition within 100 ps [Fig. \ref{nowyfig}(b)].  The intradot spin-flip transition line (the red peak near 0.4 eV in Fig. \ref{nowyfig}(a,b)) as well as fractional (1/3) resonant transition with spin flip and the charge hop (the red peak near 4.2 meV) is fully developed within 100 ps (full theory) time versus 0.5 ns  (neglected SO contribution due to  the bend).
The matrix elements for the spin-flipping transitions increase with account taken for SO effects due to the bend since the contribution of the minority spin component to both the initial
and the final states are made larger by the bend-related SO contribution [Fig. \ref{gs}]. The increased coupling between the states of opposite spin orientation
results also in the larger width of the spin-flipping transitions [cf. Fig. \ref{1scan}].

Table I presents the results also for the radii of the bend $R=15$ nm and $R=120$ nm. The spin-conserving transition rate is independent of the bend ($R$).
The spin-flipping rates decrease with $R$ -- with or without inclusion of the bend to the SO interaction. For a straight CNT the spin-flipping transitions
do not occur due to the angular orthogonality of the initial and final states unless an effect lowering the rotational symmetry is present \cite{eosprb}
(an atomic defect, a perpendicular electric field etc.).
Independent of $R$ we find a 5-fold increase of the transition rate by the contribution of the bend to the SO coupling.

\subsection{Strongly driven system}
For the larger amplitude of the AC field, $F_{AC}=4$ kV/cm [Fig. \ref{1scan}(e-h)], the evolution of the system becomes non-perturbative and the matrix elements given in Table I loose their direct relation to the actual dynamics of the quantum system.
In particular, the direct -- first order -- transition of the Rabi type  $K'\uparrow(l) \rightarrow K'\downarrow(r)$ is missing in the Figure. The width of the spin-conserving transition
$K'\uparrow(l) \rightarrow K'\uparrow(r)$ is largely increased with the amplitude of the AC field.
This spin-conserving and the spin-flipping transition $K'\uparrow(l)\rightarrow K'\downarrow(r)$ appear close
to one another on the energy scale. For large $F_{AC}$ the spin-conserving resonance extends over the frequency characteristic to the spin-flipping one. Then, a competition between the final transition states
appear and the one which prevails is the spin-conserving transition of a much larger oscillator strength [Table I].
For $F_{bias}=10$ kV/cm and $F_{AC}=4$ kV/cm the spin-flipping transitions for $\hbar\omega=12$ meV are only observed for the SO including the bend of the nanotube [Fig. \ref{1scan}(h)].
Note, that although the direct transition for the nominally resonant frequency to the right dot with the spin-flip is absent for larger $F_{bias}=2$ kV/cm and $F_{AC}=4$ kV/cm, the final state $K'\downarrow(r)$
can be reached with the second- (1/2) and third- (1/3) order transitions as given in the Fig. \ref{1scan}(e-f). For $F_{AC}=4$ kV/cm the half-resonant peaks for the transitions to $K'\downarrow(r)$ are outside the
half-resonant peak for the $K'\uparrow(r)$ transition.

Note, that the direct transition  $K'\uparrow(l)\rightarrow K'\downarrow (r)$ appears {\it faster} for a {\it larger}
bias -- see the lower panels to Fig. \ref{1scan}(a-d)  which show the minimal spin observed for simulation lasting 100, 200 and 500 ps. This is on the contrary to the behavior found for the spin-conserving $K'\uparrow(l) \rightarrow K'\uparrow(r)$ transition which slows
down for a larger bias.

\begin{figure}[h]
\begin{tabular}{ll}
\includegraphics[width=0.48\textwidth]{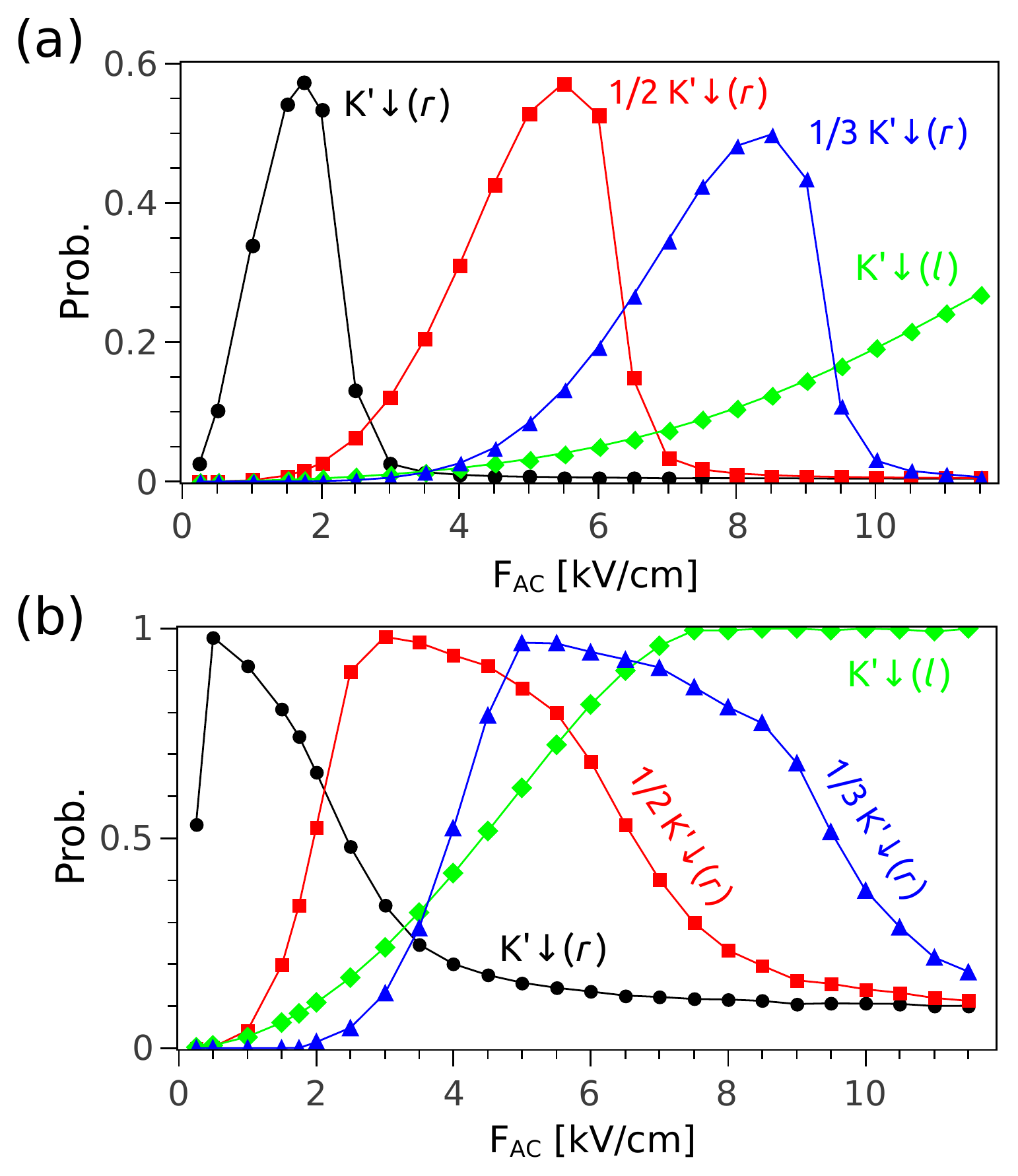}
\end{tabular}
\caption{\footnotesize \linespread{0.5} The rate of spin-flipping transitions: the maximal $|\langle\Psi(t)|K'\downarrow(l) \rangle |^2$ and $|\langle\Psi(t)|K'\downarrow(r) \rangle |^2$
obtained for 50 ps time evolution for initial state $\Psi(0) = K'\uparrow(l)$  as a function of $F_{AC}$. For each value of $F_{AC}$ the frequency $\omega$
was tuned to the resonance. A bias $F_{bias}=10$ kV/cm  was applied. Direct spin-flip transition within the left dot is plotted with the green points.
The other points correspond to spin-flip transition with charge transfer to $K'\downarrow(r)$ state realized in the direct (black color),
and fractional resonances (1/2 -- red color, and 1/3 -- blue color).
In (a) the SO interaction for a straight CNT is applied. The SO interaction that accounts for the bend is used in (b).} \label{sf}
\end{figure}

In Figure \ref{sf} we plotted the maximal probability to find the electron in spin down states obtained during 50 ps time evolution in function
of the amplitude of the AC field. For each transition the frequency was set to resonance. We note, that
{\it i)} for the spin-flipping transition within the left dot ($\rightarrow K'\downarrow(l)$) the transition rate is a monotonic slowly growing function of the AC field amplitude;
{\it ii)} the spin-flipping transitions accompanied by charge transfer from the left to the right dot is generally much faster and {\it iii)} these
transitions are non-monotonic function of the amplitude of the AC field - similar behaviour has been shown for DQDs in Ref. \onlinecite{sherman}. Here, the reason for the non-monotonic behavior is the
neighborhood of the spin-conserving transition in the frequency domain. As $F_{AC}$ increases the spin-conserving transition: both the direct
one and the fractional ones increase radically in width. Once, the spin-flipping transition finds itself within the wide peak of the spin-conserving
transition, the spin is no longer flipped as the AC field is applied: only the charge oscillation between the dots is observed.
The rate of the half-resonant transition to $K'\downarrow (r)$ state decreases for $F_{AC}>5.5$ kV/cm, when it is consumed by the half-resonant
spin-conserving transition. Then a higher transition rate can be obtained for the 1/3 resonance etc.
For the bend-related SO coupling present the maxima of the probabilites reach unity, and for lower values of $F_{AC}$.

\subsection{Transitions for the electric potentials applied along the CNT direction}
The results presented so far were obtained for the DQD potential defined along the $z$ direction.
Let us consider -- a less realistic case -- when the DQD potential and bias field are defined along the axis of the bent CNT,
by e.g.  bent gates that follow the shape of the CNT. The results for the transition matrix elements are given in Table II.
The elements for spin-flipping transitions with SO as taken for a straight CNT -- are drastically reduced as compared to Table I.
For these values the only effect triggering the spin-flip transition is the variation of the angle between the ${\bf B}$ vector
with respect to the local axis of the CNT. The bend-related SO contribution (the figures in brackets in Table II)-- breaking the rotational symmetry of the Hamiltonian -- increases the spin-flip transition rates by
as much as 40 times. The spin-conserving transition remains insensitive to the way the external electric field is introduced.


\begin{table}[htbp]
\begin{tabular}{|c|c|c|c|c|}
\hline
$R$ [nm] &$F_{bias}$& $K'\downarrow(l)$ & $K'\uparrow(r)$ & $K'\downarrow(r)$  \\ \hline
\multirow{2}{*}{30}&2 kV/cm &  0.003 (0.125) & 4.858  (4.841) & 0.009 (0.357) \\
& 10 kV/cm & 0.0008 (0.035) & 3.062 (2.977) & 0.017 (0.682)\\ \hline
\end{tabular}
\caption{\footnotesize \linespread{0.5} Same as Table I, only for the double dot potential and the bias field defined along the axis of the bend CNT and not along the global $z$ direction.
In each cell of the column we provide two figures. The one outside (inside) the parentheses corresponds to the
SO coupling introduced for a straight (bent) CNT.}
\end{table}

\subsection{CNT with a defect}

\begin{figure}[h]
\begin{tabular}{ll}
\includegraphics[width=0.75\textwidth]{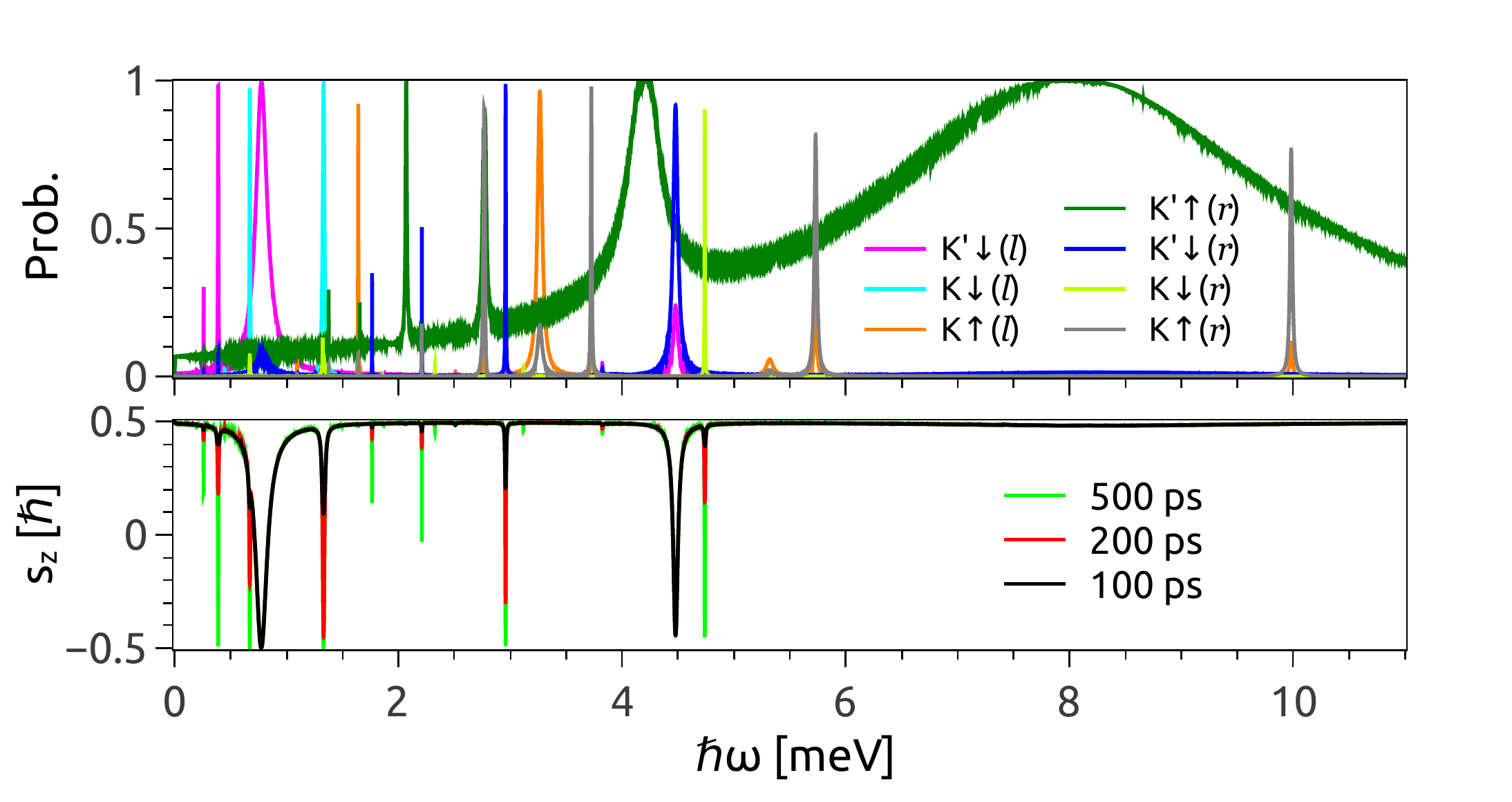}
\end{tabular}
\caption{\footnotesize \linespread{0.5} Same as figure \ref{1scan}(f) but with one carbon atom missing at a distance of 2.5 nm from the left edge of the CNT.} \label{vacant}
\end{figure}

The results presented above were obtained for a clean CNT
with a perfect crystal structure for which the inter-valley transitions are absent. In order to estimate the valley mixing effects for a CNT with crystal defects we have taken the parameters of figure \ref{1scan}(f) and removed one carbon ion at a distance of 2.5 nm from the left edge of the CNT. The results are displayed in figure \ref{vacant}. The rates of all valley-conserving transitions within $K'$ remain almost unaffected by the presence of the defect. The
inter-valley transitions which are forbidden for a clean CNT and which activated by the defect produce a sequence of narrow
peaks in the upper panel of figure \ref{vacant}. Already in figure \ref{1scan}(f) one could notice that the the peaks which corresponded to both charge and spin transitions $K'\uparrow(l)\rightarrow K'\downarrow(r)$ were accompanied by peaks--at the same frequency--for the intradot spin flips $K'\uparrow(l)\rightarrow K'\downarrow(l) $. Similar effect is observed for the 
intra-valley transitions--which occur at same frequencies for charge conserving and charge hopping processes (see the orange and
gray lines in the upper panel of figure \ref{vacant}). As far as the spin transitions are concerned the presence of the defect induces an appearance of three additional narrow lines with respect to figure \ref{1scan}(f)--one at $\hbar\omega$ = 4.7 meV with the half-resonant transition to $K\downarrow (l)$ (inverted spin, valley and dot, light green line
in figure \ref{vacant}) and another one to $K\downarrow (l)$ at left of the strong transition to $K'\downarrow (l)$ near $\hbar\omega$ = 1.4 meV and half-resonant one near $\hbar\omega$ = 0.7 meV. These two transitions occur at relatively
low rates: the lower panels of figures \ref{1scan}(f) and \ref{vacant} remain nearly
identical up to 100 ps, only after the three narrow spin-transition lines evolve in the spectrum. Concluding, for CNT with a
defect the intervalley transitions are observed. The transitions are narrow at the driving frequency scale and produce only a
slight modification to the spin flip dependence.

\section{Summary and Conclusion}
We have developed the tight-binding Hamiltonian for a carbon nanotube accounting for the spin-orbit coupling due to $\sigma$-$\pi$ hybridization
of the covalent bonds resulting from both folding the graphene sheet to a nanotube and the bend of the nanotube. For discussion of the effects
of the SO coupling contribution resulting from the bend we considered the electron transitions within a double quantum dot defined in a clean carbon nanotube in AC electric fields.
We demonstrated that although the bend-related SO interaction has a negligible influence on the energy spectra its impact on the spin-flipping transition times is pronounced.
We found that for a large amplitude of the AC voltage the spin-conserving transition peak evolves into
a wide maximum consuming the interdot Rabi oscillation for the spin-flipping transition.
We discussed the fractional resonances -- the solid state counterparts of the multiphoton transitions.
We demonstrated that for higher AC amplitude the fractional resonances can be used to perform the spin-flips with interdot charge transfer.

\section{Acknowledgements}
This work was supported by National Science Centre
according to decision DEC-2013/11/B/ST3/03837 and by PL-GRID infrastructure.

\end{document}